\def\beq{\begin{equation}}
\def\eeq{\end{equation}}

\def\eqi{\begin{equation}}
\def\eqf{\end{equation}}
\def\eqia{\begin{eqnarray}}
\def\eqfa{\end{eqnarray}}

\def\ton#1{\left(#1\right)}

\documentclass[longtitle,sort&compress]{elsarticle}
\usepackage{amssymb,amsmath}
\usepackage{epsfig}

\RequirePackage{color}

\begin{document}

\begin{frontmatter}

\title{Constraining the Schwarzschild-de Sitter Solution in Models of Modified Gravity}

\author[auth1]{Lorenzo Iorio\corref{cor}}
\ead{lorenzo.iorio@libero.it}

\author[auth2a,auth2b,auth2c]{Matteo Luca Ruggiero}
\ead{matteo.ruggiero@polito.it}

\author[auth3a,auth3b]{Ninfa Radicella}
\ead{ninfa.radicella@sa.infn.it}

\author[auth4a,auth4b]{Emmanuel N. Saridakis}
\ead{Emmanuel\_Saridakis@baylor.edu}

\cortext[cor]{Corresponding author.}

\address[auth1]{Ministero dell'Istruzione, dell'Universit\`{a} e della Ricerca (M.I.U.R.)-Istruzione, Viale Unit\`{a} di Italia 68, 70125, Bari (BA), Italy}

\address[auth2a]{Dipartimento di Fisica, Universit\`{a} di Torino, Via Pietro Giuria 1, 10125 Torino, Italy}
\address[auth2b]{DISAT, Politecnico di Torino, Corso Duca degli Abruzzi 24, 10129 Torino, Italy}
\address[auth2c]{INFN, Sezione di Torino, Via Pietro Giuria 1, 10125 Torino, Italy}

\address[auth3a]{Dipartimento di Fisica E. Caianiello, Universit\`a di Salerno, Via
Giovanni Paolo II 132,  Fisciano (Sa), Italy}
\address[auth3b]{INFN, Sezione di Napoli, Gruppo Collegato di Salerno,  Napoli, Italy}

\address[auth4a]{CASPER, Physics Department, Baylor University, Waco, TX
76798-7310, USA}
\address[auth4b]{Instituto de F\'{\i}sica, Pontificia Universidad de Cat\'olica de
Valpara\'{\i}so, Casilla 4950, Valpara\'{\i}so, Chile}

\begin{abstract}
The Schwarzschild-de Sitter (SdS) solution exists  in the large majority of modified
gravity theories, as expected, and in particular the effective cosmological constant is
determined by the specific parameters of the given theory. We explore the possibility to
use future extended radio-tracking data from the currently ongoing New Horizons mission in
the outskirts peripheries of the Solar System, at about 40 au, in order to constrain this
effective cosmological constant, and thus to impose constrain on each scenario's
parameters. We investigate some of the recently most studied modified gravities, namely
$f(R)$ and $f(T)$ theories, dRGT massive gravity, and Ho\v{r}ava-Lifshitz gravity, and we
show that New Horizons mission may bring  an improvement of one-two
orders of magnitude  with respect to the present bounds from planetary orbital dynamics.
\end{abstract}


\end{frontmatter}

\section{Introduction}\label{sec:intro}

General Relativity (GR) has undergone brilliant successes  since its inception 100
years ago (see, e.g., the review \cite{2015Univ....1...38I} and references therein).
Einstein's theory is the standard paradigm for describing the gravitational interaction,
verified by many  experimental evidences \cite{Will:2014bqa}, even though, with the
possible exception of binary-pulsar systems, at least to a certain extent, these tests
are probes of the weak-field gravity, or differently speaking they probe gravity up to
intermediate scales ($\simeq 1-10^1$ au). Nevertheless, one of the current challenges in
theoretical physics and cosmology is the description of gravitation at large scales.
In particular, evidences from astrophysics and cosmology
\cite{Perlmutter:1997zf,Riess:1998cb,Tonry:2003zg,
Knop:2003iy,
Barris:2003dq,
Riess:2004nr, Astier:2005qq,Eisenstein:2005su,Spergel:2006hy,Hinshaw:2012aka}
suggest  that the Universe content is  76\% \textrm{dark energy}, 20\% \textrm{dark
matter}, 4\% ordinary baryonic matter. This implies that in order to reconcile
the observations with GR we are led to assume that the Universe is dominated by
\textrm{dark entities}, with peculiar characteristics. The dark energy is an exotic cosmic
fluid, which has not yet been detected directly, and which does not cluster as ordinary
matter; indeed, its behaviour closely resembles that of the cosmological constant
$\Lambda$, which, in turn, brings about other problems, concerning its nature and origin
\cite{Peebles:2002gy,Martin:2012bt}. On the other hand, the dark matter is an unknown
type of matter, which
has the clustering properties of ordinary matter; since 1933 it is has been related to
the problem of missing matter in astrophysical scenarios \cite{Zwicky:1933gu}. Moreover,
some kind of cold and pressureless  dark matter (whose distribution is that of a spherical
halo around the galaxies)  is also required to explain the rotation curves of spiral
galaxies \cite{Binney87}. Hence, the best answer we have today for these cosmic puzzles
is the so called concordance model or $\Lambda$CDM,  which provides the simplest
description of the available data concerning the large-scale structure of the Universe. \textcolor{black}{For a recent review, see e.g. \cite{2016PDU....12...56B}.}
This picture is completed with the inflationary scenario which solves the horizon,
flatness and monopole problems \cite{kolb1990early}.\\
\indent Besides these difficulties in explaining observations, there are theoretical
motivations suggesting that a theory of gravity more fundamental than GR should be
formulated: Einstein's theory is not renormalizable, and thus it cannot be quantized as
is. In a recent paper by Berti et al. \cite{2015CQGra..32x3001B}, a thorough review of the motivations
to consider extensions of GR can be found together with a discussion of some modified
theories of gravity (see also the recent reviews \cite{2011PhR...509..167C,
2015Univ....1...92H, 2015Univ....1..186B, 2015Univ....1..199C, 2015Univ....1..446Z} and
references therein). \\
\indent A possible  strategy towards a new theory of gravity is, in some sense, a natural
generalization of Einstein's approach, according to which \textrm{gravity is geometry}.
Accordingly, a new theory is obtained extending GR on a purely geometric basis: in other
words, the required ingredients to match the observations or to solve the theoretical
conundrums derive from a geometric structure richer than that of GR.\\
\indent As a  prototype of this strategy, which has gained an increasing attention during
the last decade, we  mention  the $f(R)$ theories, where the gravitational Lagrangian
depends on a function  of the scalar curvature $R$; extensive reviews can be found in
 \cite{Capozziello:2007ec, 2007Geom, Sotiriou:2008rp, defelice, 2011PhR...505...59N}. These theories
are also referred to as ``extended theories of gravity'', since they naturally generalize
GR: in fact, when $f(R)=R$ the action reduces to the usual Einstein-Hilbert action, and
Einstein's theory is obtained.\\
\indent Motivations for studying these theories can be different but, as clearly
synthesized by Sotiriou and Faraoni \cite{Sotiriou:2008rp}, they can be considered as toy-theories that are
relatively simple to handle and that allow to study the effects of the deviations from
Einstein's theory with sufficient generality. For instance $f(R)$ theories provide
cosmologically viable models, where both the inflation phase and the late-time
accelerated expansion
are reproduced; furthermore, they have been used to explain the rotation curves of
galaxies without need for dark matter (see \cite{Capozziello:2007ec, Sotiriou:2008rp, defelice} and references therein). These theories can be
studied in the metric formalism, where the action is varied with respect to metric
tensor, and in the Palatini formalism, where the action is varied with respect to the
metric and the affine connection, which are supposed to be independent from one another
(there is also the metric-affine formalism, in which the matter part of the action
depends on the affine connection, and is then varied with respect to it). In general, the
two approaches are not equivalent:  the solutions of the Palatini field equations are a
subset on solutions of the metric field equations \cite{magnano}.\\
\indent A different approach to the extension of GR derives from a
generalization of Teleparallel Gravity (TEGR) \cite{pereira,Maluf:2013gaa}:
this theory is based on a Riemann-Cartan space-time, endowed with the non symmetric
Weitzenb\"ock connection which, unlike the Levi-Civita connection of GR, gives rise to
torsion but it is curvature-free. In TEGR torsion determines the geometry, while the
tetrad field is the dynamical one; the field equations are obtained from a
Lagrangian containing the torsion scalar $T$, arising from contractions of the torsion
tensor. Notwithstanding GR and TEGR have a different geometric structure, they have the
same dynamics: in other words,  every solution of GR is also solution of TEGR and vice
versa. Hence,
one could start from TEGR and extend its Lagrangian from $T$ to an arbitrary function
$f(T)$, resulting to the so-called $f(T)$ gravity \cite{Ferraro:2008ey,Linder:2010py}
(for a review see \cite{Cai:2015emx}). Since $f(T)$ gravity is different from TEGR,
$f(T)$
theories have been considered as potential candidates to describe the
cosmological behavior
\cite{cardone12,Chen:2010va,sari11,Myrzakulov:2010vz,Yang:2010hw,bengo,kazu11,
Karami:2013rda, cai11,capoz11,Bamba:2013jqa,Camera:2013bwa}. Additionally, various
aspects of $f(T)$ gravity have been considered, such as for instance, exact solutions and
stellar models
\cite{Wang:2011xf,Ferraro:2011ks,Gonzalez:2011dr,
Capozziello:2012zj,Rodrigues:2013ifa,Nashed:uja,Nashed:2015qza,
Junior:2015fya,
Bejarano:2014bca,ss3,
ss4,ss6,ss7}. \\
\indent  Another possible new theory of gravity can be obtained by a massive deformation
of GR.
Endowing graviton with
a mass is a plausible modified theory of gravity that is both phenomenologically and
theoretically intriguing. From the theoretical point of view, a small non-vanishing
graviton mass is an open issue. The idea was originally introduced in the work of Fierz and Pauli
\cite{fierz39},
who constructed a massive theory of gravity in a flat
background that is ghost\,-\,free at the linearized level. Since then, a great effort has
been put in extending the result to the nonlinear level and constructing a consistent
theory. A few years ago a covariant massive gravity model has been proposed in
\cite{derham11}. Since the linearization of the mass term breaks the gauge invariance of
GR then, in order to construct a consistent theory, non-linear terms should be tuned to
remove order by order the negative energy state in the spectrum \cite{boulware72}. The
theoretical model under investigation follows from a procedure originally outlined in
\cite{arkani03,creminelli05} and has been found not to show ghosts at least up to quartic
order in the nonlinearities \cite{derham11,hassan11}. The consequent theory exploits
several remarkable features. Indeed the graviton mass typically manifests itself on
cosmological scales at late times thus providing a natural explanation of the presently
observed accelerating phase \cite{cardone12b}. Moreover, the theory allows for exotic
solutions in which the contribution of the graviton mass affects the dynamics at early
times. It actually allows for models in which the Universe oscillates indefinitely about
an initial static state, ameliorating the fine-tuning problem suffered by the emergent
Universe scenario in GR \cite{parisi12}.

Another approach that could lead towards a formulation of a quantum theory of
gravity is the Ho\v{r}ava formulation of a model that is power-counting renormalizable
due to an anisotropic scaling of space and time \cite{horava09}. This is reminiscent
of Lifshitz scalars in condensed matter physics \cite{lifshitz41a, lifshitz41b}, hence the theory is
often referred to as the Ho\v{r}ava-Lifshitz gravity. This theory has attracted a
lot
of
attention,  due to its several remarkable features in cosmology. Unfortunately, the
original model suffers from instability, ghosts, strong coupling problems and the model
has been implemented along different lines \cite{Bogdanos:2009uj}.

\indent On the other hand, if we consider  the excellent agreement of GR with  Solar
System and binary pulsar observations, it is apparent that any modified theory of gravity
should reproduce GR at the Solar System scale, i.e. in a suitable weak-field limit. In
other words, these theories must have correct Newtonian and post-Newtonian limits and, up
to intermediate scales, the deviations from the GR predictions can be considered as
perturbations. This agreement should be obtained, for all the above gravitational
modifications. In particular, all these theories  have the same spherically symmetric
solution that describes the gravitational field around a point-like source: the
Schwarzschild-de Sitter space-time (SdS).  Interestingly enough, this is a
solution of GR field equations with a cosmological constant. However, for these modified
gravities the cosmological term is not added by hand, but   it naturally
originates from the modified Lagrangian.\\
\indent In this paper,  we assume that the SdS solution can be used to model the
gravitational field of an isolated source like the Sun, and we examine the impact that
the gravitational modifications have on the Solar System dynamics. Additionally, we
explore the possibility of constraining $\Lambda$ in the distant peripheries of
the Solar System by means of the currently ongoing spacecraft-based mission New Horizons.
For a recently proposed long-range mission aimed to test long-distance modifications of
gravity in the Solar System, see \cite{2015PhRvD..92j4048B}. \\
\indent
This work is organized as follows: In Section \ref{sec:SdS} we describe the main features
of the SdS space-time, focusing on $f(R)$ and $f(T)$ theories,
massive  gravity and Ho\v{r}ava-Lifshitz gravity. Section \ref{track} is devoted to a
preliminary exposition of the experimental constraints which might be posed by using
accurate tracking of distant man-made objects traveling to the remote outskirts of the
Solar System; the case of the New Horizons probe is considered. Finally, section
\ref{theend} summarizes our results.

\section{Schwarzschild-de Sitter space-time as a vacuum solution of modified gravities}
\label{sec:SdS}

The SdS metric (see e.g \cite{2001rsgc.book.....R})
\beq
ds^{2}=\left(1-\frac{2GM}{r}-\frac{1}{3}\Lambda r^{2}
\right)dt^{2}-\frac{1}{\left(1-\frac{2GM}{r}-\frac{1}{3}\Lambda r^{2}
\right)}dr^{2}-r^{2}d\Omega^{2} \label{eq:SdSmetric}
\eeq
where $d\Omega^{2}= d\theta^2+ \sin^2 \theta d\phi^2$, is a spherically symmetric
solution
of the Einstein field equations with cosmological constant $\Lambda$ in  vacuum, namely
\beq
R_{\mu\nu}-\frac 1 2 g_{\mu \nu}R+\Lambda g_{\mu\nu}=0,
\label{eq:einsteinLambda}
\eeq
or equivalently
\beq
R_{\mu\nu}=\Lambda g_{\mu\nu},
\label{eq:einsteinLambdabis}
\eeq
around the mass $M$. The SdS space-time has been studied in connection with the
constraints arising from Solar System data \cite{Iorio:2005vw,Kagramanova:2006ax} and
moreover focusing on the effects on gravitational lensing
\cite{Rindler:2007zz,Sereno:2008kk,Ruggiero:2007jr}. In the following subsections, we
are going to show that the metric (\ref{eq:SdSmetric}) is a solution of various
gravitational modifications, under certain considerations.

\subsection{$f(R)$ theories} \label{ssec:theofR}

Let us start by summarizing the theoretical framework of the $f(R)$ theories, in
order to obtain the field equations, both in metric and the Palatini approach (see
\cite{Capozziello:2007ec, Sotiriou:2008rp, defelice} for an exhaustive
discussion), and to show that the SdS space-time is a solution.\\
\indent The field equations can be obtained by a variational principle, starting from the
action\footnote{Let the
signature of the $4$-dimensional Lorentzian manifold $\mathcal M$ be
$(+,-,-,-)$.
furthermore,  if not otherwise stated, we use units
such that $c=1$.}
\begin{equation}
S=\frac{1}{16\pi G}\int d^{4}x \sqrt{-\mathrm{det}(g_{\mu\nu)}} f (R)+S_{M}.
\label{eq:actionf(R)}
\end{equation}
As we mentioned above, in these theories the gravitational part of the Lagrangian
is represented by a function $f(R)$ of  the scalar curvature $R$, while $S_{M}$ is the
action for the matter sector, which functionally depends on the matter fields together
with their first derivatives. In the metric formalism, $\Gamma$ is supposed to
be the Levi-Civita connection of the metric $g$ and, consequently, the scalar
curvature $R$ has to be intended as $R\equiv R(g)
=g^{\alpha\beta}R_{\alpha \beta}(g)$. On the contrary, in the
Palatini formalism the metric $g$ and the affine connection
$\Gamma$ are supposed to be independent, so that the scalar
curvature $R$ has to be intended as $R\equiv R( g,\Gamma)
=g^{\alpha\beta}R_{\alpha \beta}(\Gamma )$, where $R_{\mu \nu
}(\Gamma )$ is the Ricci-like tensor of the connection $\Gamma$.\\
\indent In the metric formalism the action (\ref{eq:actionf(R)})
is varied with respect to the metric $g$, and one obtains the
following field equations
\begin{align}
f'(R) R_{\mu \nu }-\frac{1}{2}f(R)g_{\mu \nu }-\left(\nabla _{\mu
}\nabla _{\nu }-g_{\mu \nu }\square \right)f'(R)= {8\pi
G} \,\mathcal{T}_{\mu \nu}, \label{eq:fieldmetric1}
\end{align}
where $f'(R)=df(R)/d R$,  $\nabla_{\mu}$ is the covariant derivative associated with
$\Gamma$, $\square \equiv \nabla_{\mu} \nabla^{\mu}$, and  $\displaystyle
{\mathcal{T}}^{\mu\nu} = -\frac{2}{\sqrt
g}\frac{\delta S_{M}}{\delta g_{\mu\nu}}$ is the
standard minimally coupled matter energy-momentum tensor. The
contraction of  the field equations (\ref{eq:fieldmetric1}) with
the metric tensor leads to the scalar equation
\begin{align}  \label{eq:fieldmetricscalar1}
3\square f'(R)+f'(R)R-2f(R)={8\pi G} \mathcal{T},
\end{align}
where $\mathcal{T}$ is the trace of the energy-momentum tensor. Note that Eq.
(\ref{eq:fieldmetricscalar1}) is a differential equation for the
scalar curvature $R$, while in GR the scalar curvature is
algebraically related to $\mathcal{T}$ through $\displaystyle R=-{8\pi G} \mathcal{T}$.\\
\indent In the Palatini formalism, by
independent variations with respect to the metric $g$ and the
connection $\Gamma$, we obtain the following equations of motion:
\begin{eqnarray}
f^{\prime }(R) R_{(\mu\nu)}(\Gamma)-\frac{1}{2} f(R)  g_{\mu \nu
}&=&{8\pi G}{} \mathcal{T}_{\mu \nu },  \label{ffv1}\\
\nabla _{\alpha }^{\Gamma }\textcolor{black}{(} \sqrt{g} f^\prime (R) g^{\mu \nu
})&=&0, \label{ffv2}
\end{eqnarray}
where  $\nabla^{\Gamma}$ denotes covariant derivative with respect
to the connection $\Gamma$.  Actually, it is possible to show
\cite{FFVa,FFVb} that the manifold $\mathcal{M}$, which is the model of the
space-time, can be a posteriori endowed with a bi-metric structure
$(\mathcal{M},g,h)$  equivalent to the original metric-affine structure
$(\mathcal{M},g,\Gamma)$, where $\Gamma$ is assumed to be the Levi-Civita
connection of $h$. The two metrics are conformally related by
\begin{equation}\label{h_met2}
h_{\mu \nu }=f^\prime (R)  \;  g_{\mu \nu }.
\end{equation}
The equation of motion (\ref{ffv1})
can be supplemented by the scalar-valued equation obtained by
taking the contraction of (\ref{ffv1}) with the metric tensor:
\begin{equation}
f^{\prime} (R) R-2 f(R)= {8\pi G}{} \mathcal{T}.  \label{ss}
\end{equation}
Equation (\ref{ss}) is an algebraic equation for the scalar
curvature $R$, thus slightly generalizing the GR case where $R$ is proportional
to $T$.

In order to compare the predictions of $f(R)$ gravity with Solar System dynamics data, we
have to
consider the solutions of the field equations in vacuum. As a consequence,  in the metric
approach
the field equations read
\begin{align}
f'(R) R_{\mu \nu }-\frac{1}{2}f(R)g_{\mu \nu }-\left(\nabla _{\mu
}\nabla _{\nu }-g_{\mu \nu }\square \right)f'(R)= 0,
\label{eq:fieldmetric1vac}
\end{align}
supplemented with the scalar equation
\begin{align}  \label{eq:fieldmetricscalar1vac}
3\square f'(R)+f'(R)R-2f(R)=0.
\end{align}
In the the Palatini approach, the field equations in vacuum become
\begin{eqnarray}
f^{\prime }(R) R_{(\mu\nu)}(\Gamma)-\frac{1}{2} f(R)  g_{\mu \nu
}&=&0,  \label{ffv1vac}\\
\nabla _{\alpha }^{\Gamma }\textcolor{black}{(} \sqrt{g} f^\prime (R) g^{\mu \nu
})&=&0, \label{ffv2vac}
\end{eqnarray}
and they are supplemented by the scalar equation
\begin{equation}
f^{\prime} (R) R-2 f(R)=0.
\label{ssvac}
\end{equation}

It is useful to emphasize some  features of the scalar equations
(\ref{eq:fieldmetricscalar1vac})
and (\ref{ssvac}), which
can help to understand the differences between the vacuum solutions in the two
formalisms.
In
Palatini $f(R)$ gravity,  the trace equation (\ref{ssvac}) is
an algebraic equation for $R$, which admits constant solutions
$R=c_{i}$ \cite{FFVa}, and it is identically satisfied if $f(R)$
is proportional to $R^2$. As a consequence, it is easy to verify
that (if $f'(R) \neq 0$) the field equations become
\begin{equation}
R_{\mu\nu}=\frac 1 4 R g_{\mu\nu},
\label{eq:gralambda1}
\end{equation}
which are the same as the GR field equations with a cosmological constant
(\ref{eq:einsteinLambdabis}). In particular, we now have
\begin{equation}
 \Lambda_{fR}=\frac {1}{ 4}
R.
\label{LambdafR}
\end{equation}
 In other words, in the Palatini formalism, in vacuum, we can obtain only
solutions that describe space-times with constant scalar curvature $R$.
Summarizing, Eq. (\ref{eq:gralambda1})
suggests that all GR solutions with cosmological constant  are
solutions of vacuum Palatini field equations: the function $f(R)$
only determines the  solutions of algebraic equation (\ref{ssvac}).
\\
\indent
In metric $f(R)$ gravity the trace equation
(\ref{eq:fieldmetricscalar1vac}) is a differential equation for
$R$: this implies that, in general, it admits more solutions than the
corresponding Palatini equation. In particular, we notice that if
$R=\mathrm{constant}$ we obtain the Palatini case. Hence for a given
$f(R)$ function, in vacuum,  the solutions of the field
equations of Palatini $f(R)$ gravity are a subset of the solutions
of the field equations of metric $f(R)$ gravity \cite{magnano}. However, in metric
$f(R)$ gravity vacuum solutions with variable
$R$ are allowed too (see, e.g., \cite{metricspherically}).\\
\indent Therefore, if we confine ourselves with constant scalar curvature, we have shown
that in $f(R)$ gravity the SdS space-time (\ref{eq:SdSmetric}) is a solution of the field
equations, and in particular the ``effective'' cosmological constant term depends on the
analytical expression of $f(R)$.\\
\indent As for the reliability of these solutions for describing without conceptual
drawbacks the gravitational field of a star, like the Sun,  the issue has been lively
debated in the literature (see e.g. \cite{Sotiriou:2008rp,defelice}). In the
Palatini formalism, the possibility of constructing vacuum solutions that match an
internal solution has been discussed, and it has been shown that when
one considers even a simple model such as a polytropic star, divergences arise. However,
things are different for non-analytical $f(R)$, and also the role of the conformal
metric $h_{\mu\nu}$ can help to avoid these singularities (see \cite{lorenzof2015} and
references therein).
On the other hand, metric $f(R)$ gravity is in agreement with Solar System tests only if
the chamaleon mechanism is considered, according to which the additional scalar degree of
freedom of the theory is a function of the curvature: the mass of the scalar field is
large at Solar System scale, in order not to affect the dynamics, while it is
small at cosmological scale, in order to drive the accelerated expansion. For a thorough
discussion about the reliability of $f(R)$ gravity see the reviews
\cite{Sotiriou:2008rp, defelice}, where it is discussed that Palatini $f(R)$
gravity, beyond the above mentioned difficulty with polytropic stars, suffers from other
problems, which make acceptable models practically indistinguishable from $\Lambda$CDM.
On the other hand, in metric $f(R)$ gravity it is possible to obtain models that are
in agreement with observations, having peculiarities that make it possible, at least in
principle, to distinguish them from $\Lambda$CDM. However, we are not going to get into
the details of the above debate, since for the purpose of the present work it is adequate
that $f(R)$ gravity admits the SdS solution. Finally, we recall that some properties of SdS and Reissner-Nordstr\"{o}m (SdS generalised) black holes in $f(R)$ modified gravity were investigated in \cite{2013CQGra..30l5003N, 2014PhLB..735..376N}.

\subsection{$f(T)$ theories} \label{ssec:theoT}

In this subsection we outline the theoretical framework of $f(T)$ gravity and we obtain
the
field equations that accept the SdS space-time as solution \cite{Cai:2015emx}). In $f(T)$
gravity the tetrads are the dynamical fields. Given a coordinate basis, the components
$e^a_\mu$ of the tetrads are related to the metric tensor
through $g_{\mu \nu}(x) = \eta_{a b} e^a_\mu(x) e^b_\nu(x)$, with  $\eta_{a b} =
\text{diag}(1,-1,-
1,-1)$. We point out that, in our notation,  latin indices refer to the tangent space,
while greek indices label coordinates on the manifold. The field equations can be
obtained by varying the action
\begin{equation}
S = \frac{1}{16 \pi G} \int{ f(T)\, e \, d^4x} + S_M
\label{eq:action}
\end{equation}
with respect to the tetrads, where   $e = \text{det} \  e^a_\mu =
\sqrt{-\text{det}(g_{\mu
\nu})}$
and $S_M$ is the action for the matter fields. In the action (\ref{eq:action}),
$f$ is a
differentiable function of the torsion scalar $T$: in particular, if $f(T)=T$, the action
is the
same as in TEGR, and the theory is equivalent to GR. In terms of the tetrads one defines
the torsion tensor as
\beq
T^\lambda_{\ \mu \nu} = e^\lambda_a \left( \partial_\nu e^a_\mu - \partial_\mu e^a_\nu
\right ), \
\label{eq:deftorsiont}
\eeq
and the ``super-potential'' tensor
\beq
S^\rho_{\ \mu \nu} = \frac{1}{4} \left ( T^{\rho}_{\ \ \mu \nu} - T_{\mu \nu}^{\ \
\rho}+T_{\nu \mu}
^{\ \ \rho} \right ) +
\frac{1}{2} \delta^\rho_\mu T_{\sigma \nu}^{\ \ \sigma} - \frac{1}{2} \delta^\rho_\nu
T_{\sigma \mu}
^{\ \ \sigma},  \label{eq:defcontorsion}
\eeq
from which one obtains the torsion scalar
\beq
T = S^\rho_{\ \mu \nu} T_\rho ^{\ \mu \nu}.  \label{eq:deftorsions}
\eeq
By variation of the action (\ref{eq:action}) with respect to the tetrad field $e^a_\mu$,
we obtain
the field equations
\beq
e^{-1}\partial_\mu(e\  e_a^{\ \rho}   S_{\rho}^{\ \mu\nu})f_T-e_{a}^{\ \lambda}
S_{\rho}^{\ \nu\mu}
T^{\rho}_{\ \mu\lambda} f_T
+  e_a^{\ \rho}  S_{\rho}^{\ \mu\nu}\partial_\mu (T) f_{TT}+\frac{1}{4}e_a^{\nu} f = 4\pi
G e_a^{\ \mu} {\mathcal{T}}_\mu^\nu,
\label{eq: fieldeqs}
\eeq
where  ${\mathcal{T}}^\nu_\mu$ is the matter energy-momentum tensor, and where the
subscripts $T$ denote differentiation with respect to $T$.\\
\indent We are interested in static spherically symmetric solutions that can be used to
describe
the gravitational field of a point-like source, e.g. of the Sun. To this end, we write
the
space-
time metric in the form
\begin{equation}
ds^2=e^{A(r)}dt^2-e^{B(r)}dr^2-r^2 d\Omega^2 \ . \label{metric}
\end{equation}
In the usual, ``pure-tetrad'' formulation of $f(T)$ gravity, the above metric is produced
by the non-diagonal tetrad (\cite{tamanini12,Krssak:2015oua})
\footnotesize
\begin{multline}
\!\!\!\!\!\!\!\!\!\!\!{e_\mu}^a=
\left(
\begin{array}{cccc}
 e^{A/2} & 0  & 0 & 0\\
 0 & e^{B/2} \sin \theta \cos \phi & e^{B/2} \sin \theta \sin \phi  & e^{B/2} \cos \theta
\\
 0 & -r \left(\cos \theta \cos \phi \sin \gamma+\sin \phi \cos\gamma \right) & r
\left(\cos \phi \cos\gamma -\cos \theta \sin \phi \sin\gamma \right) & r
   \sin \theta \sin\gamma \\
 0 & r \sin \theta \left(\sin \phi \sin\gamma -\cos \theta \cos \phi \cos\gamma \right) &
-r \sin \theta \left(\cos \theta \sin \phi \cos\gamma +\cos
   \phi \sin\gamma \right) & r \sin ^2\theta \cos\gamma
\end{array}
\right),
\label{eq:rotatedtetrad}
\end{multline}\normalsize

where $\theta$, $\phi$ are rotation angles, and $\gamma(r)$ is a general function of $r$.
The expression of the torsion scalar for the above tetrad turns out to be
\begin{equation}
T(r) = \frac{2\, e^{-B}}{r^2} \left[
    1 + e^B + 2\,e^{B/2} \sin\gamma +
    2\, e^{B/2}\, r\,  \gamma' \cos\gamma \\+
    r\, A' \left(1+e^{B/2}\sin\gamma\right)
  \right]\,. \label{eq:torsionscalar}
\end{equation}
We are interested in extracting static vacuum  solution with constant
torsion scalar  $T=T_{0}$  (i.e. $T'=0$). The field equations (\ref{eq: fieldeqs}) become
\begin{eqnarray}
     \frac{f_0}{4} -\frac{f_{T_0}\,e^{-B}}{4r^2}\left( 2-2\,e^B+r^2e^B T_0-2r\,B'
\right)&=&0 \label{eq:fieldqT1} \,,\\
    -\frac{f_0}{4} +\frac{f_{T_0}\,e^{-B}}{4r^2} \left( 2-2\,e^B+r^2e^B T_0-2r\,A'
\right)&=&0 \label{eq:fieldqT2} \,,\\
   4-4\,e^B -r^2A'^2 +2r\,B'+r\,A'\left(2+r\,B'\right) -2r^2A'' &=&0 \,,
\label{eq:fieldqT3}
\end{eqnarray}
where $f_{0}\equiv f(T_{0})$, $f_{T_{0}}\equiv f_{T}(T_{0})$ and prime denotes
differentiation with
respect to
$r$.  We point out that spherically symmetric solutions with non constant torsion scalar
$T' \neq
0$ have been already investigated \cite{Iorio:2012cm,Ruggiero:2015oka} and Solar System
constraints
 have been discussed \cite{Iorio:2015rla,2013MNRAS.433.3584X}.\\
\indent It is possible to show (see \cite{tamanini12}) that the unique solution of the
equations (\ref{eq:fieldqT1})-(\ref{eq:fieldqT3}) is given by
\begin{eqnarray}
&&e^{A(r)}= 1- \frac{2\,M}{r}-\frac{\Lambda_{fT}}{3}r^2,\nonumber\\
&&e^{B(r)}=e^{-A(r)},
\label{eq:solAfT}
\end{eqnarray}
with
\beq
\Lambda_{fT} = \frac{1}{2} \left(\frac{f_0}{f_{T_0}}-T_0\right) \,. \label{eq:defLambdaf}
\eeq
Thus, in the theory at hand one obtains an ``effective'' cosmological constant,
determined by the functional form of $f(T)$, and thus he obtains a SdS solution. Notice
however that, because of the presence of the arbitrary function $\gamma(r)$ in the
definition of the torsion scalar (\ref{eq:torsionscalar}), knowing $\Lambda_{fT}$ cannot
constrain $f(T)$, since an arbitrary value of $\Lambda$ can be achieved by fine tuning
$T_{0}$ with a suitable choice of $\gamma(r)$. In other words, when the
torsion tensor is constant, any $f(T)$ model admits the solution in the form  of
(\ref{eq:solAfT}), with given values of $M$ and $\Lambda_{fT}$. On the contrary, in the
case of $f(R)$ theories, the value of the scalar curvature $R$, that is proportional to
$\Lambda_{fR}$, strictly depends on the function $f(R)$, since it is obtained from the
trace
equation (\ref{ssvac}).

\subsection{Massive gravity}\label{spec:massive}

In this subsection we summarize the basic part of massive gravity formulation relevant to
the present analysis. Specifically, we are interested in static spherically symmetric
solutions in which the mass term becomes identical to the cosmological constant term.\\
\indent The possibility of endowing graviton with a mass goes back to 1939, where Fierz
and Pauli constructed the linearized theory of non-interacting massive gravitons in a
flat background \cite{fierz39}. Unfortunately, the solutions of this theory do not
continuously connect with those of GR in the limit of zero graviton mass,
and this is the famous van Dam, Veltman and Sakharov (vDVZ) discontinuity
\cite{vanDam70, zakharov70}. This  vDVZ discontinuity can be alleviated at the nonlinear
level through the
Vainshtein mechanism \cite{vainshtein72}, however these nonlinearities produce
the so-called Boulware-Deser (BD) ghost degree of freedom \cite{boulware72}.\\
\indent In 2010 a ghost-free theory was proposed by de Rham, Gabadadze and Tolley (dRGT)
\cite{derham11}. In the standard formalism of dRGT theory, the dynamics is determined by
a modified action written in terms of a dynamical metric $g_{\mu\nu}$ and an arbitrary
fiducial metric $f_{\mu\nu}$ needed to construct the gravitational self-interacting
potential $\mathcal{U}$. The corresponding action reads:
\begin{equation}\label{totaction}
S=-\frac{1}{8\pi G}\int \left(\frac{1}{2} R+m^2  \mathcal{U}\right) \sqrt{-g} \  d^{4}x \
+ \ S_{M},
\end{equation}
where   $S_M$ describes ordinary matter which is supposed to directly interact only with
$g_{\mu\nu}$. The potential term, coupled through the graviton mass $m$, is defined by
\cite{nieuwenhuizen11}
\begin{align}\label{potential}
\mathcal{U}\nonumber &= \frac{1}{2} (K_1^2-K_2)+\frac{c_3}{6}(K_1^3 -3 K_1K_2+2K_3)+\\ \nonumber \\
& + \frac{c_4}{12}(K_1^4 -6K_1^2 K_ 2+3 K_2^2 +8K_1 K_3 -6 K_4),
\end{align}
with
$K_n$ denoting the traces of a tensor $K^{\mu}_{\nu}$ constructed from the inverse metric
$g^{\mu\nu}$ and the fiducial one through
$$
K^\mu_\nu=\delta^\mu_\nu-\sqrt{g^{\mu\rho} f_{ab}\partial_\rho \phi^a\partial_\nu \phi^b},
$$
and  with $K_n\equiv \text{tr} K^n$.
The four fields $\phi^a$ are the St\"uckelberg fields\footnote{St\"uckelberg fields were
originally
introduced by St\"uckelberg in 1938 to restore gauge-invariance in electromagnetism but
the method
works
equivalently well for spin-2 fields.}, which transform as scalars under coordinate
transformations,
such that the fixed metric $f_{\mu\nu}$
$$
f_{\mu\nu}=f_{ab}\partial_\mu \phi^a\partial_\nu \phi^b,
$$
as well as the quantity $g^{\mu\alpha}f_{\alpha\nu}$, are promoted to tensor fields,
while the potential $U(g,f)$ to a scalar.
Potential (\ref{potential}) has been shown to be the most general potential for a
ghost-free theory of massive gravity in four dimensions \cite{hassan11}.\\
\indent Apart from interesting cosmological features, the dRGT massive gravity
admits the SdS solution where the ``effective'' cosmological
constants arises due to the graviton mass. In particular, considering the choice
$$
c_4=1+c_3+c_3^2
$$
in (\ref{potential}), then the mass term of the theory behaves exactly as the
cosmological constant term in GR for a spherically symmetric ansatz
\cite{berezhiani12}, and the resulting expression for the metric reads as follows:
\begin{equation}
ds^2=-\left(1-\frac{2 G M}{r} -\frac{\Lambda}{3} r^2\right)\  dt^2+\frac{1}{1-\frac{2 G
M}{r} -\frac{\Lambda}{3} r^2} \ dr^2+r^2\  d\Omega^2,
\end{equation}
that is the standard SdS solution of GR in static
coordinates.
The difference here is that it is accompanied by nontrivial background of the
St\"uckelberg fields.
In terms of the parameters of the theory the effective cosmological constant reads
\begin{equation}\label{lambdamassive}
\Lambda_{mg}=\frac{2m^2}{1+c_3}.
\end{equation}
Finally, note that this solution allows to recover GR when $c_3+c_4>0$
\cite{koyama11}, below the so-called Vainshtein radius $r_V=\left(GM/m^2\right)^{1/3}$.
\textcolor{black}{For extended dRGT models, see, e.g., \cite{2015PhRvD..92l4063T, 2016arXiv160104399W}.}

\subsection{Ho\v{r}ava-Lifshitz gravity}\label{spec:horava lifshitz}

Let us summarize Ho\v{r}ava-Lifshitz gravity \cite{horava09}, in order to extract its
spherically symmetric solutions. As we mentioned in the Introduction, Ho\v{r}ava-Lifshitz
gravity is a power-counting renormalizable theory, obtained through an anisotropic
scaling of space and time in the Ultraviolet limit. This feature allows for the inclusion
of higher-dimensional spatial derivative operators that dominate in the high energy
limit, while in the Infrared lower-dimensional operators take over, presumably
providing a healthy low-energy limit, namely GR. Additionally, the absence of higher
order time derivative terms prevents ghost instabilities. However, as it becomes obvious,
the anisotropic scaling breaks Lorentz invariance, and breaking of general covariance has
been shown to introduce a dynamical scalar mode that may lead to strong coupling problem
and instabilities \cite{Bogdanos:2009uj,wang11}.\\
\indent Recently, a new covariant version of Ho\v{r}ava Lifshitz gravity has been
formulated by Ho\v{r}ava and Melby-Thompson \cite{HMT10} in which, in order to heal the
scalar graviton problem, two auxiliary scalar fields have been introduced: the Newtonian
pre-potential $\phi(t,x)$ and the gauge field $A(t,x)$. The latter eliminates the new
scalar degree of freedom, thus curing the strong coupling problem in the Infrared limit,
and general covariance is restored. In the following we refer to the covariant version of
Ho\v{r}ava and Melby-Thompson, and the running coupling $\lambda$  in
the extrinsic curvature term of the action is not set to $1$  \cite{daSilva11}.\\
\indent
With the perspective of Lorentz symmetry breaking, the suitable variables in
Ho\v{r}ava-Lifshitz theory are the lapse function, the shift vector and the spatial
metric,
$N$, $N_i$, $g_{ij}$ respectively, according to the Hamiltonian formulation of General
Relativity developed by Dirac \cite{Dirac58} and Arnowitt, Deser and Misner \cite{adm59}.
Then the line element can be rewritten as
$$
ds^2=-N^2 dt^2+g_{ij} \left(dx^i+N^i dt\right)\left(dx^j+N^j dt\right).
$$
The theory can be assumed to satisfy the projectability condition, i.e. the lapse
function only depends on time
$N=N(t)$, while the total gravitational action is given by
\begin{equation}\label{HLaction}
S_g=\zeta^2\int dt\  d^3x\
N\sqrt{g}\left(\mathcal{L}_K-\mathcal{L}_V+\mathcal{L}_\phi+\mathcal{L}_
A+\mathcal{L}_{\lambda}\right),
\end{equation}
where $g=\text{det}(g_{ij})$ and
\begin{eqnarray*}
\mathcal{L}_K&=&K_{ij} K^{ij} -\lambda K^2,\\
\mathcal{L}_\phi&=&\phi\ \mathcal{G}^{ij}\left(2K_{ij}+\nabla_i\nabla_j \phi\right),\\
\mathcal{L}_A&=&\frac{A}{N}\left(2 \Lambda_g-R\right),\\
\mathcal{L}_\lambda&=&(1-\lambda)\left[(\nabla \phi)^2+2 K \nabla^2\phi\right].
\end{eqnarray*}
Note that  in this subsection covariant derivatives and Ricci terms refer to the
$3$-metric $g_{ij}$. $K_{ij}$ represents the extrinsic curvature
$$
K_{ij}=g_i^k\nabla_k n_j,
$$
$n_j$ being a unit normal vector of the spatial hypersurface,  and $\mathcal{G}_{ij}$ is
the $3$-~dimensional generalised Einstein tensor
$$
\mathcal{G}_{ij}=R_{ij}-\frac{1}{2}g_{ij} R+\Lambda_g g_{ij}.
$$
 We mention that the parameter $\lambda$ characterizes deviations of the kinetic part of
the action from GR. The most general parity-invariant Lagrangian density
up to six order in spatial derivatives reads as \cite{sotiriou09}
\begin{align}
\mathcal{L}_{V} \nonumber &= 2 \zeta^2 g_0 +g_1R+\frac{1}{\zeta^2}\left(g_2\  R^2+g_3\  R_{ij}
R^{ij}\right)+\\ \nonumber \\
\nonumber & + \frac{1}{\zeta^4}\left[g_4\  R^3+g_5\  R R_{ij} R^{ij}+g_6\  R^i_j R^j_k R_i^{k} + \right. \\ \nonumber \\
& +\left. g_7\  R \nabla^2 R+ g_8\  (\nabla_i R_{jk}) (\nabla^i R^{jk})\right],
\end{align}
where in physical units $\zeta^2=(16 \pi G)^{-1}$,  $G$ being the Newtonian constant, and
the couplings $g_s\  (s=0,1,\dots,8)$ are all dimensionless.\\
\indent
We are here interested in vacuum static spherically symmetric solutions. These have been
derived in \cite{greenwald10} and \cite{lin12} for the case $\lambda=1$ and
$\lambda\neq1$, respectively (see also \cite{Kiritsis:2009sh,Kehagias:2009is}). Omitting
the details of the derivation, and despite the
large class of solutions, we mention that in both cases the SdS
solution can be extracted. Hence, various constraints on the parameters and functions of
the theory are derived, both due to equations of motion and Solar System tests
\cite{greenwald10,lin12}. When $\lambda=1$, which is the GR value, and similarly to what
happens in the original version presented in \cite{HMT10}, the SdS
solution is recovered, with the choice $\phi=A=0$, and the effective cosmological
constant arises from the $g_0$ coupling, namely
\begin{equation}\label{LambdaHL}
\Lambda_{HL}=\frac{1}{2}\zeta^2 g_0.
\end{equation}
On the other hand, if one desires to consider $\lambda$ as a free parameter, one has to
consider the Newtonian pre-potential $\phi$, as well as the gauge field $A$, as part of
the metric on which matter fields couple, as shown in \cite{lin14}.

\section{Preliminary sensitivity analysis on the possibility of
constraining $\Lambda$ with New Horizons}\label{track}

\subsection{Suggested data analysis}

\begin{figure}[ht]
\centering
\centerline{
\vbox{
\begin{tabular}{cc}
\epsfysize= 8.0 cm\epsfbox{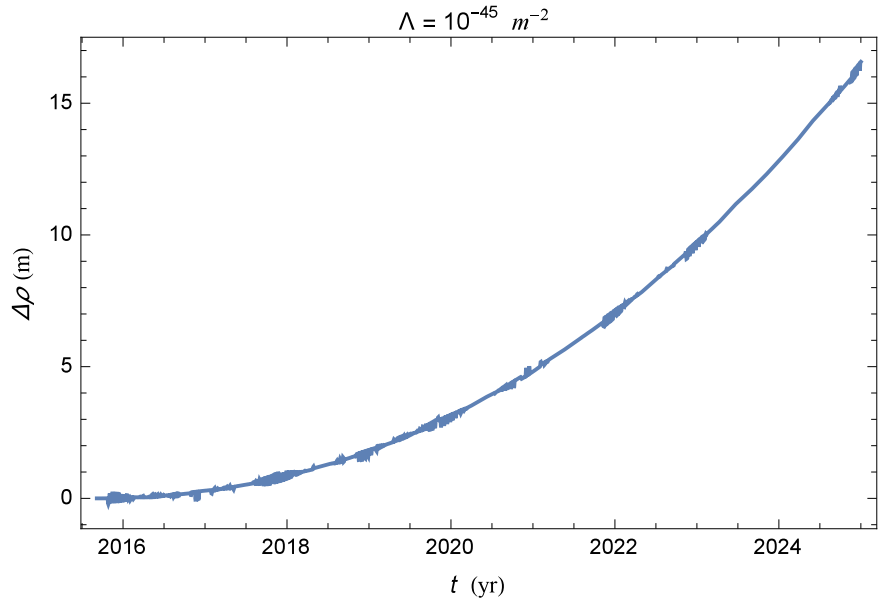} \\
\end{tabular}
}}
\caption{Simulated signature $\Delta\rho$ induced by $\Lambda = 10^{-45}$ m$^{-2}$ on the
geocentric range of New Horizons over a decade-time span 2015-2025. It was obtained by
taking the difference \textcolor{black}{$\Delta\rho(t)$} between  \textcolor{black}{two time series of $\rho(t)=\sqrt{\ton{x_{\textrm{NH}}(t) - x_\oplus(t)}^2 + \ton{y_{\textrm{NH}}(t) - y_\oplus(t)}^2 + \ton{z_{\textrm{NH}}(t) - z_\oplus(t)}^2}$ calculated by numerically integrating the barycentric equations of motion of New Horizons and the major bodies of the Solar System in Cartesian rectangular coordinates with and
without the $\Lambda-$induced acceleration. All the standard Newton-Einstein dynamics for pointlike bodies was modeled in both the integrations which shared also the same initial conditions for August 5, 2015, retrieved from the WEB interface HORIZONS maintained by JPL, NASA. The range-rate signature $\Delta\dot\rho(t)$, not shown here, was obtained by numerically differentiating the time series for $\Delta\rho(t)$.}}
\label{figura}
\end{figure}

New Horizons \cite{2008SSRv..140....3S, 2008SSRv..140...49G} is a spacecraft which,
launched in 2006, flew by Pluto on the 14th of July 2015 without entering into orbit
around it. Orbital maneuvers were recently implemented\footnote{See
http://pluto.jhuapl.edu/News-
Center/News-Article.php?page=20151105 on the Internet.} to target the spacecraft towards
the Trans-Neptunian Object (TNO) $2014~\textrm{MU}_{69}$ of the Kuiper Belt in an extended
mission scenario.
New Horizons is spin-stabilized and therefore it will be possible to perform
radio-science experiments \cite{2010LRR....13....4T} due to the dedicated Radio
Science Experiment (REX) apparatus  \cite{2008SSRv..140..217T} carried on board and the
innovative regenerative tracking technique
\cite{2013aero.confE.170J}. The precision in Doppler measurements will be better than
$\sigma_{\dot\rho}
 = 0.1$ mm s$^{-1}$ throughout the entire mission \cite{2008SSRv..140...23F}, while
ranging will be precisely better than $\sigma_{\rho} = 10$ m (1$\sigma$) over 6 years
after 2015, i.e. at geocentric distances to beyond 50 au \cite{2008SSRv..140...23F}.\\
\indent
It is interesting to preliminarily investigate the potential ability of New Horizon's
tracking to improve the currently existing bounds on, e.g., the cosmological constant
$\Lambda$. To this aim, we will numerically simulate the range and range-rate signatures
of the extra-acceleration caused by a cosmological constant in the Solar System, by
comparing their magnitudes with the previously quoted figures for New Horizons. However,
 it should be stressed that it is just a preliminary sensitivity analysis based on the
expected precision of the probe's measurements: actual overall accuracy will be finally
set by several sources of systematic uncertainties like, e.g., the heat dissipation from
the Radioisotope Thermoelectric Generator (RTG)  and the ability in
accurately modeling the orbital maneuvers. In this respect, the extensive modeling of
such non-gravitational perturbations for the Pioneer spacecraft, recently made in the
framework of the Pioneer Anomaly investigations  \cite{
2008PhRvD..78j3001B, 2010AcAau..66..467R, 2010SSRv..151..123R, 2010SSRv..151...75B,
2011AnP...523..439R, 2012JSpRo..49..212S, 2012PhRvL.108x1101T, 2012PhLB..711..337F,
2014PhRvD..90b2004M} should
be helpful.
\\
\indent
We numerically integrate the barycentric equations of motion of the major bodies of the
Solar System and of New Horizons, with and without $\Lambda$, in Cartesian rectangular
coordinates. Both integrations share the same initial conditions, retrieved from the
WEB interface HORIZONS run by JPL, NASA, and the time interval is set to 10 yr starting
from a date posterior to the flyby of Pluto. Then, from the solutions of the perturbed and
unperturbed equations of motion, we numerically produce differential time series
$\Delta\rho(t),\Delta\dot\rho(t)$ of the Earth-New Horizons range $\rho$ and range-rate
$\dot \rho$. The amplitudes of such simulated signatures can be compared to
$\sigma_{\rho},\sigma_{\dot\rho}$ in order to preliminarily guess the value of $\Lambda$
which makes them compatible. It turns out that the range allows for tighter constraints
than the range-rate. \\
\indent
In Fig. \ref{figura} we present our results. In particular, we
depict the simulated time series $\Delta\rho(t)$ for $\Lambda =
10^{-45}$ m$^{-2}$.
%
%
It can be noticed that the size of the $\Lambda$-induced signatures is about 20 m. Thus,
the possibility of constraining $\Lambda$ to a $\simeq 10^{-45}$ m$^{-2}$ level over the
next ten years by means of New Horizons does not seem implausible. If indeed it will be
realized practically, it would represent an improvement by more than one-two orders of
magnitude with respect to the latest results appeared in the literature
\textcolor{black}{\cite{2013MNRAS.433.3584X,2014RAA....14..527L}}.
 However, it must be stressed once again that the analysis presented here has to be
intended  as a sketchy one just to explore the potential opportunity offered by New
Horizons; suffice it to say that it assumes a straightforward path over the years, without
accounting for orbital maneuvers and corrections.

\textcolor{black}{Finally, it should be remarked that the present analysis is based only on the orbital dynamics of both the major bodies of the Solar System and the probe itself. In fact, range and range-rate are not directly observable since they are calculated through the  actually measured round-trip time of flight of the photons  and their frequency shift, respectively. Thus, in principle, the impact of $\Lambda$ on the propagation of the electromagnetic waves connecting the spacecraft and the Earth \cite{2008PhRvD..77d3004S, 2008A&A...484..103S, 2012PhRvD..85f7302C, 2016PhRvD..93d4013D, 2016Univ....2....5A} should be taken into account as well. A detailed calculation of such an aspect of the measurement modeling is beyond the scopes of the present work.}

\subsection{Induced constraints on the models}

Having elaborated the observational constraints on the cosmological constant $\Lambda$
we may proceed to the constraining of the various gravitational modifications. In
particular, we will use the SdS solution and the expression
of the obtained effective $\Lambda$ in terms of the model parameters of each case,
extracted in Section \ref{sec:SdS}, in order to provide constraints and bounds on these
model parameters.
\\
\indent
 In case of $f(R)$ gravity, from the expression of the effective cosmological constant
$\Lambda_{fR}$ of (\ref{LambdafR}) we obtain a constraint on the  curvature scalar $R$
that turns
out to be constant both in metric and Palatini approach in order to have a SdS solution,
and, from
numerical estimation of $\Lambda$, we obtain  $R\sim 10^{-46}~\textrm{m}^{-2}$.
Moreover, since through the scalar
equation (\ref{eq:fieldmetricscalar1vac}) the Ricci scalar is related to the analytical
expression of the Lagrangian,  or at least to the ratio $f(R)/f^{\prime}(R)$, and thus on
the   parameters of the specific model, we can easily extracts the constraints on
them too.
 \\
\indent
In case of $f(T)$ gravity, as already remarked, the function $\gamma(r)$ can be chosen to
achieve the desired constant value of the torsion scalar through (\ref{eq:torsionscalar}),
thus the expression (\ref{eq:defLambdaf}) for $\Lambda_{fT}$ does not allow to break
this degeneracy and impose constraints on the Lagrangian.
\\
\indent
In case of massive gravity, the effective $\Lambda_{mg}$ (\ref{lambdamassive}) allows to
infer upper limits on the graviton mass. Assuming $c_3 \sim O(1)$, numerical values on
$\Lambda_{mg}$ will directly constraint $m$. Restoring SI units, i.e. replacing it with
$m_g=\hbar m/c$, the observational constraints on the cosmological constant translate
into
$$
m_g\sim10^{-69}~\textrm{g}=0.56\times10^{-36} \textrm{eV~c}^{-2}.
$$
We stress here that, as expected, our Solar System analysis can infer more stringent
constraints on the graviton mass than the analysis of of the same model using cosmological
data  \cite{cardone12b}), in which $m$ is related to the present value of the
Hubble parameter. Moreover, we can then compare our result with the upper limit $m_g <
7.68 \cdot 10^{-55}$ g from the dynamics in the Solar System \cite{talmadge88} and the
more stringent limit, namely $m_g <10^{-59}~\textrm{g}$, derived by requiring the
dynamical
properties of a galactic disk to be consistent with observations \cite{alves10} (see also
\cite{Goldhaber08} for a comprehensive review on the phenomenology of
graviton mass and experimental limits). The improvement in the obtained bounds is
obvious.
\\
\indent
Finally, for the case of Ho\v{r}ava-Lifshitz gravity, using the expression
(\ref{LambdaHL}) for the effective cosmological constant $\Lambda_{HL}$  in terms of the
coupling constant associated with the $0-th$ order spatial derivative, namely
$g_0$, we can extract its corresponding bound. It proves more
convenient to rescale $g_0$ through the Planck mass (or equivalently the
gravitational constant $\zeta^2=(16 \pi G)^{-1}$) in order to obtain a dimensionless
quantity $\tilde{g}_0$. Hence, we finally obtain
$$\tilde{g}_0\sim
10^{-113}.
$$
Similarly to the case of massive gravity, the above bound is more strict than the
corresponding cosmological ones \cite{Dutta:2009jn}.

\section{Summary and conclusions}\label{theend}

In this work we have considered that the gravitational field of an isolated source like
the Sun, can be described by the Schwarzschild-de Sitter (SdS) geometry. Such solution
exists in the large majority of modified gravity theories, as expected, and in particular
the effective cosmological constant is determined by the specific parameters of the given
theory. Hence, one can use Solar System data in order to constrain the SdS solution, and
thus eventually to extract constraints on the parameters of the gravitational
modification.\\
\indent
We have considered some of the recently most studied modified gravities, namely
$f(R)$ and $f(T)$ theories, dRGT massive gravity, and Ho\v{r}ava-Lifshitz gravity, and
after giving their SdS solution we have explored the possibility of using future extended
radio-tracking data from the currently ongoing New Horizons mission in the outskirts
peripheries of the Solar System, in order to   constrain the effective cosmological
constant, and thus the modified gravity parameters. In particular, we showed that  an
improvement of one-two orders of
magnitude may be possible, provided that steady trajectory arcs several years long will
be processed, and orbital maneuvers will be accurately modeled. Despite its necessarily
tentative and incomplete character, it turns out that such an idea should be worth of
further and more detailed consideration, especially concerning the model-building of
gravitational modifications.

\bibliography{darkbib}{}

\end{document}